\documentclass[journal]{IEEEtran}
 
\usepackage{graphicx}
\usepackage{booktabs}
\usepackage{subfigure}
\usepackage{overpic}
\usepackage{amsmath}
\usepackage{flushend}
\usepackage{amssymb}
\usepackage{multirow}
\usepackage{makecell}
\usepackage{color, soul}

\soulregister\cite7

\usepackage[ruled]{algorithm2e}
\usepackage{hyperref}
\hypersetup{
	colorlinks=true,
	linkcolor=blue,
	urlcolor=blue,
	citecolor=blue,
}

\usepackage{colortbl}
\usepackage[dvipsnames]{xcolor}
\definecolor{Bg}{HTML}{e0f1ff}

\begin{document}

\title{Convolution and Attention Mixer for Synthetic Aperture Radar Image Change Detection}
\author{Haopeng Zhang, Zijing Lin, Feng Gao, Junyu Dong, Qian Du, and Heng-Chao Li
\thanks{This work was supported in part by the National Key Research and Development Program of China under Grant 2022ZD0117202, and in part by the Natural Science Foundation of Qingdao under Grant 23-2-1-222-ZYYD-JCH.

Haopeng Zhang, Zijing Lin, Feng Gao, and Junyu Dong are with the School of Computer Science and Technology, Ocean University of China, Qingdao 266100, China. \emph{(Corresponding author: Feng Gao )}

Qian Du is with the Department of Electrical and Computer Engineering, Mississippi State University, Starkville, MS 39762 USA.

H. -C. Li is with the Sichuan Provincial Key Laboratory of Information Coding and Transmission, Southwest Jiaotong University, Chengdu 610031, China.}}

\markboth{IEEE GEOSCIENCE AND REMOTE SENSING LETTERS}
{Shell}

\maketitle

\begin{abstract}

Synthetic aperture radar (SAR) image change detection is a critical task and has received increasing attentions in the remote sensing community. However, existing SAR change detection methods are mainly based on convolutional neural networks (CNNs), with limited consideration of global attention mechanism. In this letter, we explore Transformer-like architecture for SAR change detection to incorporate global attention. To this end, we propose a convolution and attention mixer (CAMixer). First, to compensate the inductive bias for Transformer, we combine self-attention with shift convolution in a parallel way. The parallel design effectively captures the global semantic information via the self-attention and performs local feature extraction through shift convolution simultaneously. Second, we adopt a gating mechanism in the feed-forward network to enhance the non-linear feature transformation. The gating mechanism is formulated as the element-wise multiplication of two parallel linear layers. Important features can be highlighted, leading to high-quality representations against speckle noise. Extensive experiments conducted on three SAR datasets verify the superior performance of the proposed CAMixer. The source codes will be publicly available at \url{https://github.com/summitgao/CAMixer}.

\end{abstract}


\IEEEpeerreviewmaketitle

\section{Introduction}

\IEEEPARstart{S}{ynthetic} aperture radar (SAR) image change detection is widely acknowledged as a fundamental task in interpreting and understanding remote sensing data. 
It has significant implications for various applications, including land cover monitoring such as land cover monitoring \cite{vinholi22tgrs} \cite{lv22tgrs}, and disaster monitoring \cite{liu22tgrs} \cite{nava22grsl} \cite{feng22jstars}. With the increasing availability of multitemporal SAR images, the development of reliable change detection methods applicable to real-world scenarios has become crucial \cite{meng22grsl}. 

While many supervised and unsupervised methods have been proposed for SAR change detection, supervised methods often require prior knowledge and high-quality labeled samples, which are inconvenient or even difficult to collect in real applications. Furthermore, existing unsupervised methods are commonly based on convolutional neural networks (CNNs), and have limitations in long-range feature modeling. Therefore, in this letter, we primarily focus on developing robust unsupervised SAR change detection method.

Recently, Liu et al. \cite{liu19lrcnn} introduced a spatial constraint on CNN. This spatial constraint restricts the convolution operations to local regions, thereby improving change detection performance. Saha et al. \cite{saha22jstars} proposed a Siamese convolutional network. This network employs a shared set of weights to handle multi-temporal SAR images. Wang et al. \cite{9736680} employed a dual-path denoising network for SAR change detection. The network refines noise labels in training samples. Hafner et al. \cite{9570476} employed a dual-stream U-Net and performed data fusion of Sentinel-1 and Sentinel-2 images. The fusion of multi-source data, along with the dual-stream architecture, enables accurate urban change detection. Liu et al. \cite{9497508} proposed a change detection approach based on image translation. By transforming images of different types, it effectively detects changes from multi-source data, providing a versatile solution for unsupervised change detection.

Due to the inherent inductive bias in CNNs, existing methods possess the capability to discern subtle changes, such as edges and corners. Hence, the aforementioned CNN-based methods have demonstrated remarkable performance. However, with the emergence of Vision Transformer (ViT) \cite{dos21vit}, Transformer-based models have achieved significant success in various computer vision and image understanding tasks. These models utilize a global attention mechanism to capture long-range dependencies and compute informative features. Swin Transformer \cite{Liu2021SwinTH} achieves excellent performance in many vision tasks via shifted window self-attention computation. Despite their success, Transformers are rarely applied to multi-temporal SAR image analysis. Therefore, in this letter, we aim to investigate the potential of attention mechanism for SAR change detection task.

It is commonly non-trivial to design a robust Transformer-like framework for SAR change detection, since it possess the following challenges: 1) Transformers lack the inherent inductive bias of CNNs, making them less effective when training data is limited. 2) The non-linear transformation of the feed-forward network (FFN) has limitations in robust feature representation and is vulnerable to speckle noise.

\begin{figure*}[htbp]
    \centering
    \includegraphics[width=5.5in]{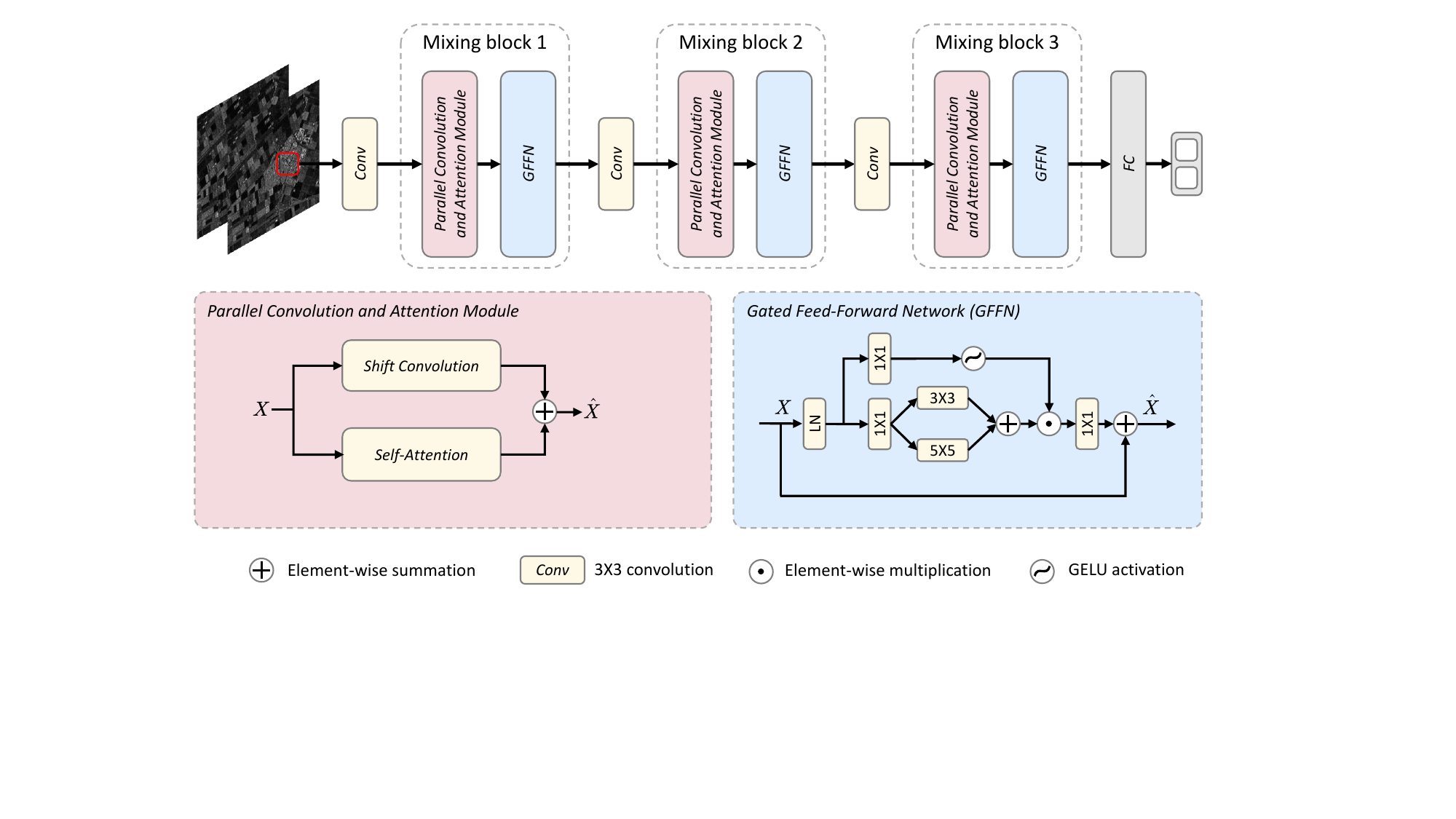}
    \caption{Illustration of the proposed Convolution and Attention Mixer (CAMixer) for SAR image change detection. The overall pipeline of CAMixer consists of three 3x3 convolutions and three mixing blocks. Each mixing block comprises a parallel Convolution and Attention Module (PCAM) and a Gated Feed-Forward Network (GFFN).}
    \label{fig:frame}
\end{figure*}

To address these challenges, we present a \textbf{C}onvolution and \textbf{A}ttention \textbf{Mixer} for SAR change detection, \textbf{CAMixer} for short. First, to compensate the inductive bias for Transformer, we combine self-attention with shift convolution in a parallel way. The parallel design enriches feature representations by modeling convolution and attention simultaneously. Additionally, we adopt a gating mechanism in FFN to enhance the non-linear feature representations. The gating mechanism is formulated as element-wise multiplication of two parallel linear layers. Important features can be highlighted, leading to high-quality representations against the speckle noise.

In a nutshell, we summarize our contributions in threefold:

\begin{itemize}

\item We present a convolution and self-attention mixed network for SAR change detection. To the best of our knowledge, we are the first to explore the Transformer-like network for multi-temporal SAR data interpretation.  

\item We propose a gated feed-forward network (GFFN) for non-linear feature transformation. Gating mechanism is formulated as the element-wise product of two parallel paths of linear transformation layers, one of which is activated with the GELU activation. Hence, the GFFN selectively emphasizes important features, thereby mitigating the interference caused by speckle noise.

\item Extensive experiments conducted on three SAR datasets demonstrate the effectiveness of the proposed CAMixer. In order to benefit other researchers, we have made our code publicly available.

\end{itemize}

\section{Methodology}

\subsection{Framework of the Proposed CAMixer}

SAR change detection aims to identify the changes that occur in the same area at different times ($t_1$ and $t_2$). The overview of CAMixer is shown in Fig. \ref{fig:frame}. 

Preclassification is performed to generate training samples for CAMixer. Specifically, we first compute the difference image by the log-ratio operator. Then, hierarchical fuzzy $c$-means \cite{gao16grsl} are used to classify the difference image into changed, unchanged, and intermediate class. The pixels from changed and unchanged class are selected as training samples.

In the proposed CAMixer, several mixing blocks are employed for local and global feature extraction. Finally, the extracted features are reshaped for classification. We now describe the key components of the mixing block: 1) Parallel Convolution and Attention Module (PCAM) and 2) Gated Feed-Forward Network (GFFN).

\begin{figure}[htpb]
    \centering
    \includegraphics[width=3.4in]{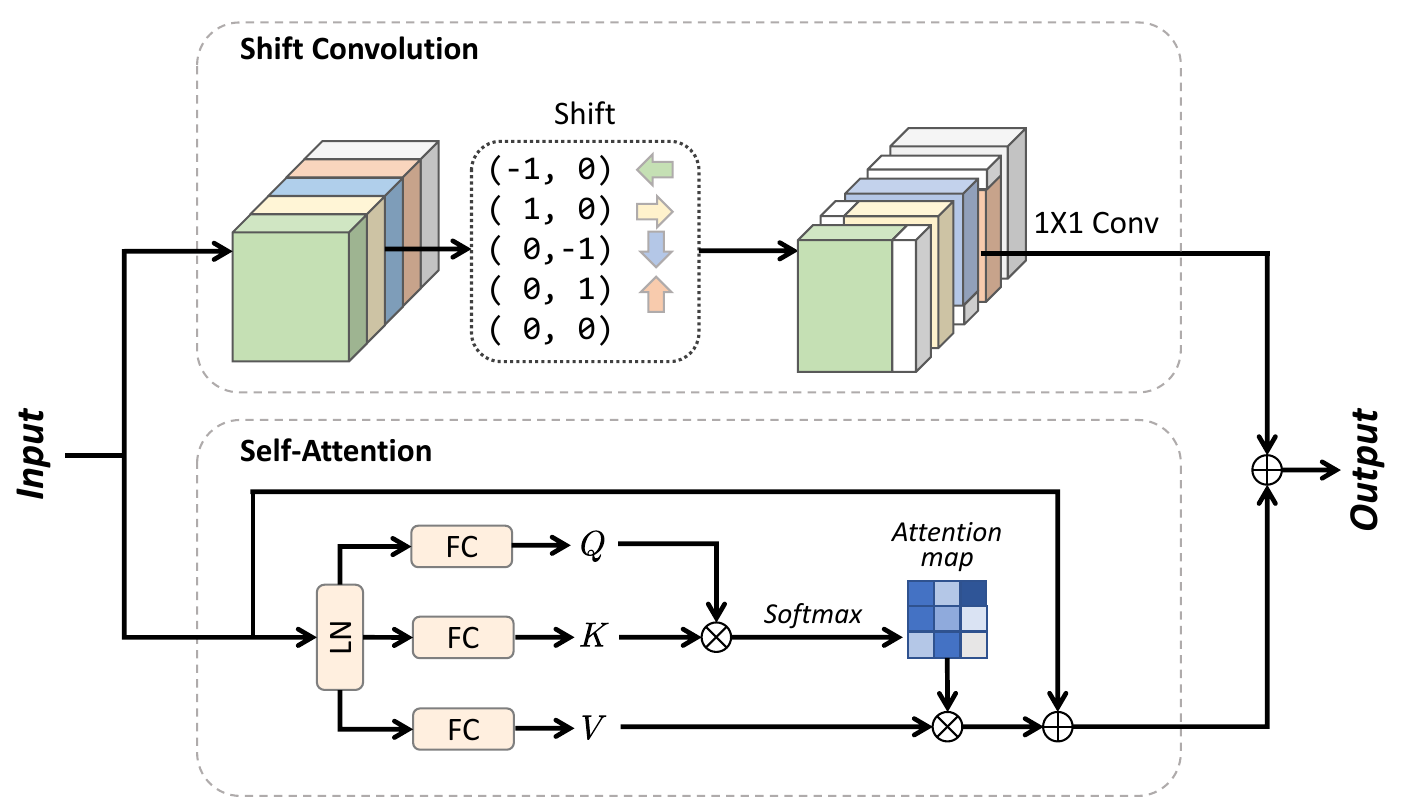}
    \caption{Details of the PCAM. It consists of shift convolution and self-attention. The output of shift convolution and self-attention are fused through element-wise summation.}
    \label{fig:pcam}
\end{figure}

\subsection{Parallel Convolution and Attention Module (PCAM)}

As shown in Fig. \ref{fig:frame}, our PCAM is composed of shift convolution and self-attention.

\textbf{Shift convolution.} Inspired by Wang's work \cite{wang2022shift}, we incorporate shift convolution for local feature extraction. It consists of a series of shift operations and a $1\times1$ convolution. The input features are evenly divided into five groups. The first four groups are shifted in different directions (left, right, top, bottom), while the last group remains unchanged.

In our implementation, we initially expand the number of channels of the input data $X$ to $\beta C$ using a $1\times1$ convolution, where $\beta$ is the expansion ratio and $C$ is the number of channels. Following the shift operation, we reduce the feature dimension back to the original size through another $1\times1$ convolution. This ensures consistency between the input and output feature sizes. Consequently, the shift convolution can be formulated as:

\begin{equation}
    \hat{X} = W^2_{1\times1}(\textrm{shift}(W^1_{1\times1}(X))),
\end{equation}
where $W^1_{1\times1}$ is the first $1\times1$ convolution, and $W^2_{1\times1}$ is the second $1\times1$ convolution. Through the shift operation, channels of the input data are shifted, enabling cross-channel information fusion through channel mixing. The second $1\times1$ convolution leverages information from neighboring pixels, while the shift convolution facilitates the incorporation of large receptive fields, while maintaining a low computational burden.

\textbf{Self-Attention Computation.} Inspired by ViT \cite{dos21vit}, we first divided the image into non-overlapping patches ($3\times3$ pixels), and encode each patch into a token embedding. Next, we compute query ($Q$), key ($K$), and value ($V$) via linear transformation of the token embedding. The output of self-attention is calculated by:
\begin{equation}
    \textrm{Attention}(Q, K, V) = 
    \textrm{Softmax}(QK^T / \sqrt{d})V,
\end{equation}
where $\sqrt{d}$ is a scaling factor. Finally, the output of shift convolution and self-attention are fused via element-wise summation. The obtained features are then normalized and fed into the GFFN to generate the input of the next mixing block.

\begin{figure}
  \centering
  \includegraphics[width=3.2in]{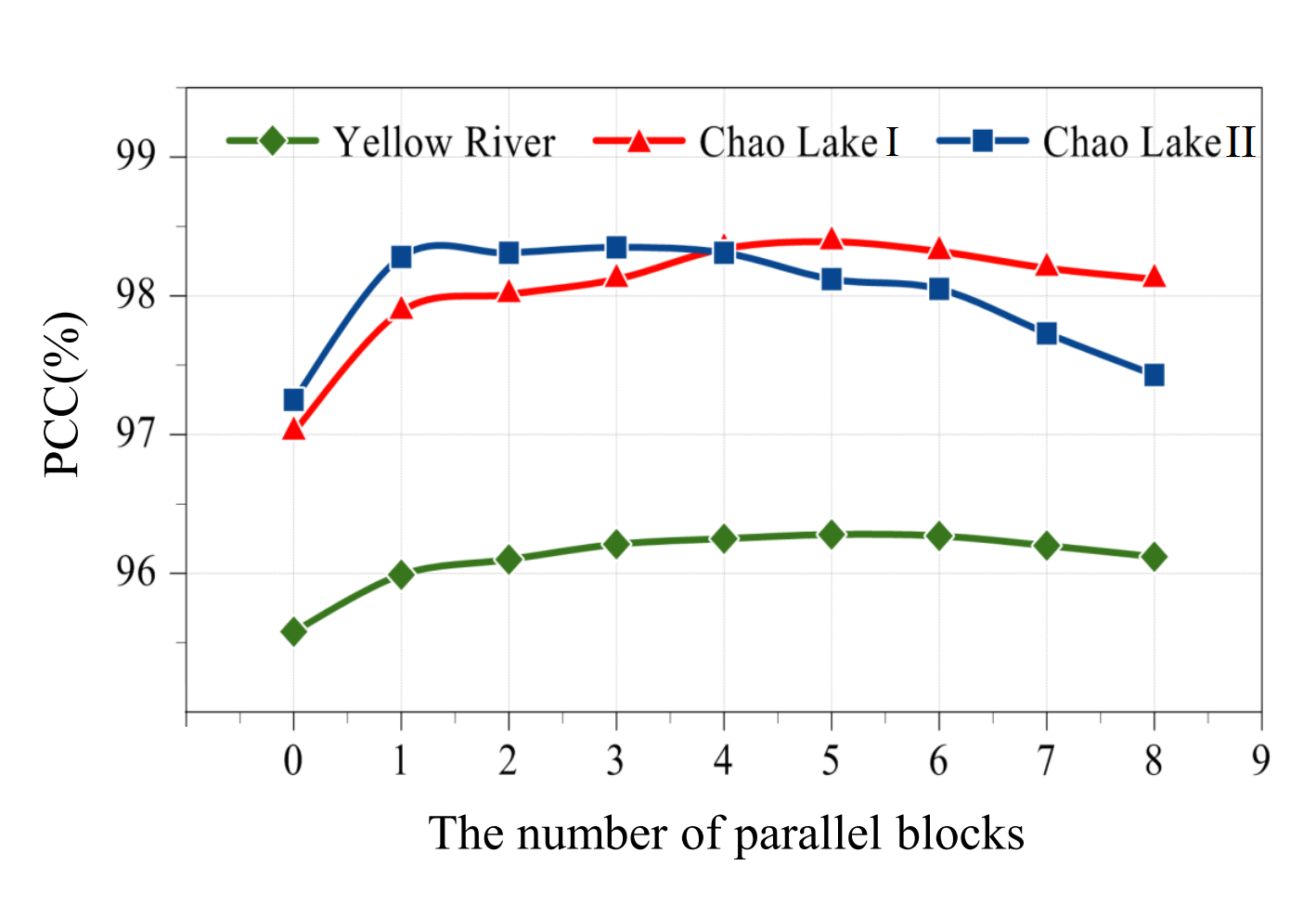}
  \caption{Relationship between the number of the parallel blocks and the PCC value.}
  \label{fig_patchsize}
\end{figure}

\begin{figure}
  \centering
  \includegraphics[width= 3.2in]{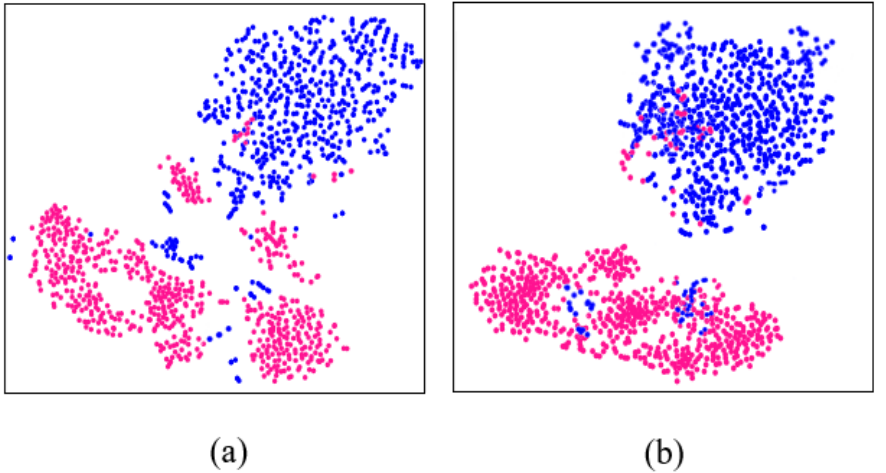}
  \caption{Visualization of the feature representations on the Chao Lake I dataset. (a) Features before the PCAM. (b) Features after the PCAM.}
  \label{fig_visual}
\end{figure}

\subsection{Gated Feed-Forward Network}

To enhance non-linear feature transformation, FFN is commonly used to process the output from the attention layer, enabling a better fit for the input of the subsequent attention layer. As illustrated in Fig. \ref{fig:frame}, we introduce the GFFN to further enhance representation learning. We make two modifications to the FFN: 1) multi-scale convolution and 2) gating mechanism. Firstly, we employ $3\times3$ and $5\times5$ depth-wise convolutions to enhance the extraction of multi-scale information. Additionally, we utilize the gating mechanism to emphasize the important components of the multi-scale convolutions.

The proposed GFFN is formulated as:
\begin{equation}\label{CGU}
    \hat{X} = W^0_{1\times1}\textrm{Gating}(X) + X,
\end{equation}
\begin{equation}\label{gating}
    \textrm{Gating}(X) =\sigma(W^1_{1\times1}(X))\odot\phi(X),
\end{equation}
\begin{equation}\label{phi}
   \phi(X)=W_{3\times3}(W^2_{1\times1}(X))+W_{5\times5}(W^2_{1\times1}(X)),
\end{equation}
where $W^0_{1\times1}, W^1_{1\times1}$, and $W^2_{1\times1}$ are $1\times1$ convolution. $W_{3\times3}$ denotes $3\times3$ depth-wise convolution, and $W_{5\times5}$ denotes $5\times5$ depth-wise convolution. Here, the $\odot$ is element-wise multiplication, and $\sigma$ is the GeLU activation. To improve computational efficiency, we reduce the expansion ratio to 2 with marginal performance loss.

\section{Experimental Results and Analysis}

\subsection{Datasets and Evaluation Metrics}

We conducted experiments on three datasets, namely the Yellow River, Chao Lake I, and Chao Lake II datasets, to validate the effectiveness of the proposed CAMixer. The Yellow River dataset covers the Yellow River Estuary region in China, with images captured in June 2008 and June 2009 using the Radarset-2 SAR sensor. The Chao Lake I and II datasets cover a region of Chao Lake in China, with images captured in May 2020 and July 2020, respectively, using the Sentinel-1 sensor. During this period, Chao Lake experienced its highest recorded water level. The ground truth change maps for all three datasets were meticulously annotated by experts with prior knowledge.

To evaluate the performance of change detection, we employ five evaluation metrics: false positives (FP), false negatives (FN), overall error (OE), percentage of correct classification (PCC), and Kappa coefficient (KC).

\subsection{Analysis of the Parallel Block Number}

There are $N$ PCAMs in the proposed CAMixer, and it is a critical parameter that may affect the change detection performance. To investigate the relationship between $N$ and change detection accuracy, we set $N$ from 0 to 8. Fig. \ref{fig_patchsize} shows that when the number of PCAM increases, the value of PCC first increases and then becomes stable. However, more PCAM would increse the computational burden. Therefore, we set $N = 3$ for the Chao Lake II dataset, and $N = 5$ for the Yellow River and Chao Lake I datasets.

\begin{figure*}[htbp]
  \centering
  \includegraphics[width=7in]{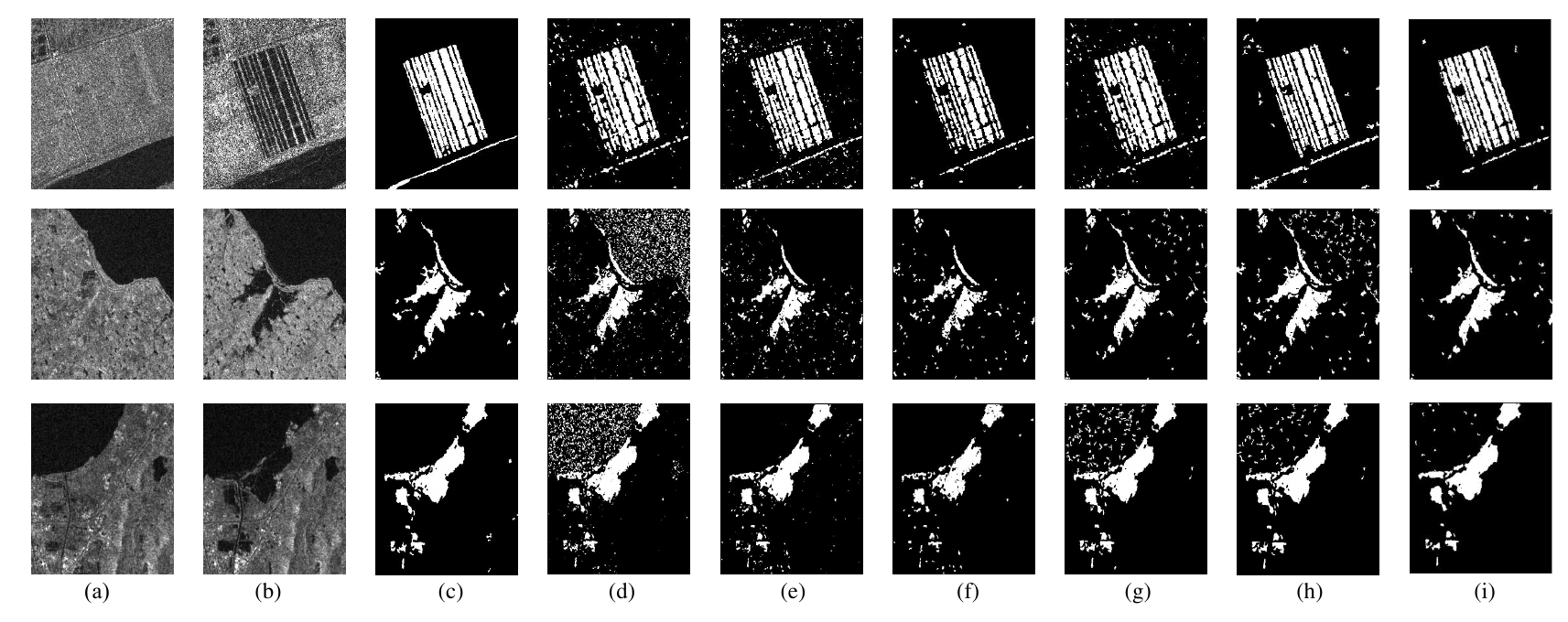}
  \caption{Visualized results of different change detection methods on the three dataset: (a) Image captured at $t_1$. (b) Image captured at $t_2$. (c) Ground truth image. (d) Result by PCA-KM \cite{celik2009unsupervised}. (e) Result by NR-CR \cite{gao2018sea}. (f) Result by NR-ELM \cite{gao2016change}. (g) Result by DDNet \cite{qu21grsl}. (h) Result by MSAPNet \cite{wang2020sar}. (i) Result by the proposed CAMixer.}
  \label{fig_result}
\end{figure*}

\begin{table}[ht]
\centering
\caption{Ablation Studies of the Proposed CAMixer}
\label{table_ablation}
\begin{tabular}{c|c c c} 
\toprule
\multirow{2}{*}{Method} 
    & \multicolumn{3}{c}{PCC on different datasets ($\%$)} \\ \cmidrule{2-4}
& Yellow River & Chao Lake1 &  Chao Lake2\\ 
\midrule
Basic Network & 95.28 & 96.40  &97.10   \\
w/o PCAM & 96.02  & 97.25  & 97.98    \\  
w/o GFFN & 96.14 & 97.58  & 98.13    \\ 
w/o H-Clustering & 96.18 & 97.76  & 98.21    \\ 
\rowcolor{Bg} Ours & 96.28 & 98.39 & 98.35  \\  
\bottomrule
\end{tabular}
\end{table}

\subsection{Ablation Study}

We conduct ablation experiments to verify the effectiveness of the PCAM and GFFN for the change detection task. We design the following four variants: 1) \textit{Basic Network} represents the backbone without PCAM and GFFN. (2) \textit{w/o PCAM} denotes the proposed method without PCAM, (3) \textit{w/o GFFN} denotes the proposed method without GFFN, and (4) \textit{w/o H-Clustering} denotes the proposed method employs fuzzy $c$-means for preclassification instead of hierarchical clustering \cite{gao16grsl}.

The results in Table \ref{table_ablation} demonstrate that compared to our full model, either \textit{w/o PCAM} or \textit{w/o GFFN} consistently exhibited lower performance on all datasets. This indicates that the PCAM significantly enhances the change detection performance, while the GFFN marginally improves it. It shows that GFFN enhances the non-linear feature transformation. Furthermore, the proposed method using hierarchical clustering demonstrates superior performance compared to \textit{w/o H-Clustering}. It is apparent that hierarchical clustering generates more reliable training samples for the proposed CAMixer, consequently enhancing the change detection performance.

To further verify the validity of our proposed PCAM, we used the t-SNE \cite{tsne} tool to visualize the characteristics before and after the module. As shown in Fig. \ref{fig_visual}, the feature representations after PCAM are noticeably more discriminative.

\begin{table}[h]
    \centering
	\caption{Change Detection Results of Different Methods on Three Datasets}
	\label{table_res}
    \begin{tabular}{c|c c c c c} 
    \toprule
    \multirow{2}{*}{Method} & \multicolumn{5}{c}{Results on the Yellow River dataset} \\
    \cmidrule{2-6}
      & FP & FN & OE & PCC ($\%$)  & KC ($\%$)\\ 
    \midrule
     PCA-KM \cite{celik2009unsupervised} & 1835 & 2798 & 4633 & 93.76 & 78.34\\
     NR-CR \cite{gao2018sea} & 2257 & 2344 & 4601 & 93.80 & 79.03\\ 
     NR-ELM \cite{gao2016change} & 629 & 3806 & 4435 & 94.03 & 77.80\\
     DDNet \cite{qu21grsl} & 1239 & 2161 & 3400 & 95.42 & 84.12\\
     MSAPNet \cite{wang2020sar} & 1206 & 2026 & 3232 & 95.65 & 84.96\\
     \rowcolor{Bg} Proposed CAMixer & 619 & 2145 & 2764 & 96.28 & 86.86\\
    \bottomrule
    \multicolumn{6}{c}{}  \\
    \toprule
    \multirow{2}{*}{Method} 
    & \multicolumn{5}{c}{Results on the Chao Lake I dataset}\\
    \cmidrule{2-6}
      & FP & FN & OE & PCC ($\%$)  & KC ($\%$)\\ 
    \midrule
     PCA-KM \cite{celik2009unsupervised} & 12126  & 1786 & 13912 & 90.57 & 52.08\\
     NR-CR \cite{gao2018sea} & 2906 & 2892 & 5798 & 96.07 & 71.34\\ 
     NR-ELM \cite{gao2016change} & 2282 & 3370 & 5652 & 96.17 & 70.71\\
     DDNet \cite{qu21grsl} & 3858 & 1182 & 5040 & 96.58 & 77.60\\
     MSAPNet \cite{wang2020sar} & 6632 & 1108 & 7740 & 94.75 & 68.95\\
     \rowcolor{Bg} Proposed CAMixer & 1178 & 1203 & 2381 & 98.39 & 89.84\\
    \bottomrule
    \multicolumn{6}{c}{}  \\
    \toprule
    \multirow{2}{*}{Method} 
    & \multicolumn{5}{c}{Results on the Chao Lake II dataset} \\ 
    \cmidrule{2-6}
      & FP & FN & OE & PCC ($\%$)  & KC ($\%$)\\ 
    \midrule
     PCA-KM \cite{celik2009unsupervised} & 8432 & 2273 & 10705 & 92.74 & 65.69\\
     NR-CR \cite{gao2018sea} & 959 & 2397 & 3356 & 97.72 & 86.63\\
     NR-ELM \cite{gao2016change} & 595 & 3836 & 4431 & 97.00 & 81.27\\
     DDNet \cite{qu21grsl} & 3107 & 779 & 3886 & 97.36 & 86.18\\
     MSAPNet \cite{wang2020sar} & 2006 & 837 & 2843 & 98.07 & 89.55\\
     \rowcolor{Bg} Proposed CAMiser & 1416 & 1019 & 2435 & 98.35 & 90.84\\
    \bottomrule
    \end{tabular}
\end{table} 

\subsection{Experimental Results and Comparison}

We compare the proposed CAMixer with five baselines, including PCA-KM \cite{celik2009unsupervised}, NR-CR\cite{gao2018sea}, NR-ELM\cite{gao2016change}, DDNet \cite{qu21grsl} and MSAPNet \cite{wang2020sar}. Fig. \ref{fig_result} illustrates the visual comparison of the change maps generated by different methods on three datasets. The corresponding quantitative evaluation metrics are illustrated in Table \ref{table_res}.

\textit{Results on the Yellow River dataset:} The Yellow River dataset is severely degraded by speckle noise. As a result, it is difficult to obtain satisfactory results by conventional methods. The qualitative results of this dataset are shown in the first row of Fig. \ref{fig_result}. It can be observed that the proposed CAMixer suppresses the false alarms effectively, and the change map of CAMixer is the most similar to the ground truth. Furthermore, the CAMixer reports the best PCC value, gaining 0.63\% and 0.86\% improvement of PCC over DDNet and MSAPNet, respectively. DDNet and MSAPNet are CNN-based methods, and it is evident that CAMixer improves the change detection performance by introducing the Transformer-like architecture. 

\textit{Results on the Chao Lake I and II datasets:} The qualitative results of Chao Lake I and II datasets are shown in the second and third rows of Fig. \ref{fig_result}, respectively. The proposed CAMixer greatly reduces the false alarms, and obtains the best PCC and KC values on both datasets. It is evident that the proposed CAMixer improves the feature representations via parallel convolution and self-attention computation. The parallel design of shift convolution and self-attention extracts local and global features simultaneously, leading to high-quality representations against the speckle noise. Additionally, the GFFN selectively emphasizes critical features, which further mitigates the interference caused by speckle noise.

From the above experiments on three SAR datasets, it can be seen that the proposed CAMixer has better performance than several traditional methods and CNN-based methods. Furthermore, CAMixer reports the best KC values, gaining 1.90\%, 12.24\%, and 1.29\% improvement over the second-best one on three datasets, respectively. It should be noted that KC value is of the most convincing evaluation metrics for SAR change detection. Moreover, the CAMixer obtains balanced FP and FN values on three datasets. It demonstrates that the PCAM captures abundant convolution and self-attention feature interactions, and contributes to better change detection results.

\section{Conclusions and Future Work}

In this letter, we propose CAMixer, a novel SAR change detection network that produces reliable change detection results. To address the inductive bias limitation of Transformer-like networks, we combine self-attention with shift convolution in a parallel manner. Moreover, we propose a gated feed-forward network to enhance non-linear feature transformation, formulated as the element-wise multiplication of two parallel linear layers. Extensive experiments on three SAR change detection datasets demonstrate the superiority of CAMixer and validate the effectiveness of its two critical components. In the future, we plan to investigate the fusion of mutli-source remote sensing data to improve the change detection performance.

\bibliography{re} 
\bibliographystyle{IEEEtran}

\end{document}